\documentclass[12pt]{article}
\usepackage{amsmath,amssymb}
\usepackage{graphics}
\usepackage[dvipdfm]{graphicx}
\usepackage{color}
\usepackage{url}

\makeatletter

\@addtoreset{equation}{section}
\makeatother

\begin{document}
\begin{titlepage}
\begin{flushright}
{\small EPHOU-12-005}\\[-1mm]
{\small OU-HET 754/2012}
\end{flushright}
\vspace*{1.2cm}

\begin{center}

{\Large\bf 
Enhancement of Higgs to diphoton decay width \\
in non-perturbative Higgs model
} 
\lineskip .75em
\vskip 1.5cm

\normalsize
{\large Naoyuki Haba}$^1$, 
{\large Kunio Kaneta}$^{1,3}$,
{\large Yukihiro Mimura}$^2$,
and
{\large Ryo Takahashi}$^{1}$

\vspace{1cm}

$^1${\it Department of Physics, Faculty of Science,
 Hokkaido University, Sapporo, 060-0810, 
 Japan} \\

$^2${\it Department of Physics, 
 National Taiwan University, 
 Taipei, 10617, Taiwan (R.O.C.)} \\

$^3${\it Department of Physics, Graduate School of Science,
 Osaka University, Toyonaka, Osaka 560-0043, 
 Japan} \\

\vspace*{10mm}

{\bf Abstract}\\[5mm]
{\parbox{13cm}{\hspace{5mm}
%
We investigate a possibility
if a loop diagram via Higgsino can enhance the Higgs to diphoton decay width
in supersymmetric models with an extension of Higgs sector.
A model with an additional non-renormalizable term of Higgs fields
is firstly analyzed where the higher order term can introduce the Higgs coupling
to Higgsinos as well as charged Higgs bosons.
We point out that a choice of the Higgs coupling to obtain a significant size
of enhancement of diphoton decay width reduces the Higgs mass
and/or a size of non-renormalizable term needs to be large
and a cutoff scale is around the weak scale.
Another model in which the Higgsino mass term is generated by
a non-perturbative instanton effect via a strong dynamics in a context of SUSY QCD
is also suggested.
It is shown that
the sign of the Higgs coupling to fermions is opposite from perturbative models
due to an operator including bosonic fields in the denominator
and
a constructive contribution to the diphoton decay amplitude
can be easily obtained in this kind of model.
}}

\end{center}

\end{titlepage}


\baselineskip=18pt

\section{Introduction}

The CMS/ATLAS collaborations at the Large Hadron Collider (LHC)
released a historical announcement about the observation 
of a new particle consistent with a Higgs boson \cite{Englert:1964et,Glashow:1961tr} at about 5 sigma~\cite{CMS-ATLAS}.
Needless to say, it is important to test that the observed boson
is really identified to the Higgs boson in the standard model (SM) or not.
The current data analyses agree with the SM prediction.
Possible hint of the deviation from the SM prediction is an excess of
the $h \to \gamma\gamma$ channel, especially at the ATLAS experiment~\cite{CMS-ATLAS}\footnote{The ATLAS experiment has reported $\sigma/\sigma_{\rm SM}=1.80\pm0.30(\text{stat})^{+0.21}_{-0.15}(\text{syst})^{+0.20}_{-0.14}\text{(theory)}$~\cite{ATLAS}.}:
\begin{eqnarray}
\sigma/\sigma_{\rm SM} = 1.56 \pm 0.43  \quad ({\rm CMS} ), \\
\sigma/\sigma_{\rm SM} = 1.9 \pm 0.5  \quad ({\rm ATLAS} ) .
\end{eqnarray}
More statistics will be needed to determine if the excess is real or
just due to a statistical fluctuation, in the experimental side.
In the theoretical side, simultaneously, it is worth to investigate the possibility
of the enhancement of the diphoton partial decay width, without significantly modifying the total
decay width or production cross section of the Higgs boson in SM \cite{ArkaniHamed:2012kq}.


The Higgs coupling to photons is induced at the loop-level \cite{Ellis:1975ap,Shifman:1979eb},
and therefore, it is sensitive to the presence of new charged particles
which couple to the Higgs boson \cite{Carena:2012xa}.
The colored particle can also modify the production cross section via gluon fusion,
as well as the partial digluon decay width.
Such modification via the colored particle may become significant rather than
the diphoton rate due to a color factor.
Therefore, if only the diphoton decay width differs from the SM expectation,
a colorless charged particle is a preferable target.
In SM, the dominant contribution comes from a $W$ boson loop,
and a correction comes from the top quark loop, which gives a 
destructive contribution to the $W$ boson loop.
In order to enhance the diphoton rate, one needs a constructive contribution
to the $W$ contribution.
The loop contributions via sequential chiral fermions whose masses are generated by 
Higgs vacuum expectation values (VEVs)
always generate destructive ones,
and thus, a devised structure of the couplings to the Higgs boson
is needed. 
The LEP experiments provide a strong bound of the mass of light new charged particles,
and as a consequence,
a large coupling to the Higgs boson is implied if the diphoton decay width is significantly modified.

The mass of the boson observed at the LHC is 125--126 GeV.
Such mass of the Higgs boson may require a new physics, such as supersymmetry (SUSY),
if a stabilization condition is applied to the Higgs self-coupling \cite{Cabibbo:1979ay,EliasMiro:2011aa} (see also \cite{Djouadi:2005gi} and references therein).
Minimal SUSY standard model (MSSM) is, of course, nicely compatible with the 
125 GeV Higgs boson \cite{Okada:1990vk,Ellis:1990nz,Haber:1990aw}.
The enhancement of diphoton rate can be also realized within 
MSSM \cite{Carena:2011aa,Hagiwara:2012mg,Buckley:2012em}
because there are plenty of new charged particles.
However, a large loop correction is needed to the Higgs boson mass
because the Higgs boson mass is predicted to be around the $Z$ boson mass at the tree level.
The required large loop correction needs a heavy stop mass or large $A_t$ trilinear scalar term
and they may spoil naturalness which is one of the motivation of SUSY \cite{Giardino:2012ww}.
Extension of the Higgs sector in MSSM is also the issue to explain the 125 GeV mass naturally (e.g. see \cite{Arbey:2011ab,Carmi:2012yp} also for its various implications). 

In this paper, we consider an extension of the Higgs sector in a SUSY model,
and investigate whether the enhancement of diphoton decay width and
increasing the Higgs mass can cooperate in the model.
The simplest extension of MSSM may be so-called Next-to-MSSM \cite{Ellis:1988er}
in which a singlet field
which couples to the Higgs fields is added.
We will concentrate a possibility
of non-renormalizable term in the Higgs sector
without adding a new field.
Those extensions of MSSM are related to the origin of the Higgsino mass (so-called $\mu$),
which is indeed one of the issues in MSSM.
We consider a model in which the Higgsino mass 
depends on the Higgs VEV, 
and see if a large Higgs coupling to Higgsino (SUSY partner of the Higgs boson)
can be obtained to enhance the diphoton decay width.

This paper is organized as follows: In section 2, we give a brief review of the Higgs to diphoton decay width. In section 3, a possibility of enhancement of the diphoton decay width and a realization of the Higgs mass with 125 GeV are investigated in a SUSY model with non-renormalizable Higgs term, namely Beyond MSSM (BMSSM). The discussions of diphoton decay width enhancement with the appropriate Higgs mass is presented also in a non-perturbative Higgs model in section 4. Section 5 is devoted to our conclusion and discussion.
%

\section{The Higgs to diphoton decay width}

The Higgs to diphoton decay is obtained loop diagram,
and the analytical expression can be found in literature \cite{Ellis:1975ap,Shifman:1979eb,Djouadi:2005gi}.
In SM, the leading contribution comes from the $W$ boson loop
and the next-to-leading contribution is from the top quark loop.
The expression of the diphoton partial decay width can be found in \cite{Ellis:1975ap,Shifman:1979eb}.
A convenient formula to study the diphoton decay width
in terms of the Higgs coupling to the charged particles in the loop is
given in \cite{Carena:2012xa}:
\begin{equation}
\Gamma(h\to \gamma\gamma)
= \frac{\alpha^2 m_h^3}{1024 \pi^3}
\left|
\frac{g_{hVV}}{m_V^2} Q_V^2 A_1(\tau_V)
+
\frac{2 g_{hf\bar f}}{m_f} N_{c,f} Q_f^2 A_{1/2} (\tau_f)
+ 
N_{c,S} Q_S^2 \frac{g_{hSS}}{m_S^2} A_0 (\tau_S)
\right|^2.
\end{equation}
In the expression, 
$V$, $f$, and $S$ refer to generic vector, fermion and scalar particles, respectively,
$Q_i$ is the electric charge of the particle,
and $N_{c,i}$ is the number of particles with color.
The loop functions $A_{1,1/2,0}$ are found in the references.
If $\tau_i \equiv 4m_i^2/m_h^2 >1$ (no on-shell decays to the charged particles), 
the loop function $A_1 (\tau_V)$
is negative, and $A_{1/2} (\tau_f)$ and $A_0(\tau_S)$ are positive.
The SM contributions come from $W$ boson ($V=W$) and top quark ($f=t$),
and the quantities of the loop functions in the case of $m_h = 125$ GeV are
\begin{equation}
A_1 = -8.32, \qquad N_c Q_t^2 A_{1/2} = 1.84.
\end{equation}
The couplings of $hWW$ and $ht\bar t$ in SM are obtained as
$g_{hWW}/m_W^2 = 2 g_{ht\bar t}/m_t = 2/v$, where
$v \approx 246\ {\rm GeV}$.

The enhancement of diphoton decay width via fermion or scalar loop
can be obtained if $g_{hf\bar f}/m_f <0$ or $g_{hSS}<0$.
If the new charged fermion is a chiral fermion such as sequential fourth generation,
its loop contribution is always destructive to the SM contribution because
$g_{h f\bar f}/m_f$ has same sign of $g_{hWW}$ coupling (which is positive in the current convention).

Surely, it is possible to have a constructive contribution via fermion loop
if there are VEV-independent Dirac mass terms as well as the Yukawa interaction \cite{Carena:2012xa}.
%
%
In the next section, 
we will study if the Higgsino loop can enhance the diphoton rate
by a constructive contribution to the $W$ boson loop.

\section{Non-renormalizable Higgs term (BMSSM)}

In MSSM there exists Higgs-Higgsino-gaugino coupling,
and therefore,
chargino loop diagram can contribute to the diphoton decay amplitude.
However, the interaction is the $SU(2)_L$ weak gauge coupling,
and the charginos have to be heavier than about 100 GeV due to the LEP bound.
As a consequence, the loop diagram via a gaugino-like chargino that mainly consists of gaugino
cannot provide a significant contribution
to the decay amplitude.
Hereafter, we neglect the gaugino-Higgsino mixing for simplicity to describe,
and we consider the contribution from the Higgs-Higgsino-Higgsino coupling
and the loop diagram via a chargino which mainly consists of Higgsino.

The Higgs-Higgsino-Higgsino coupling can be generated
if we add a non-renormalizable term to the superpotential:
\begin{equation}
W = \mu H_u \cdot H_d + \frac{c}{\Lambda} (H_u \cdot H_d)^2.
\end{equation}
This type of superpotential is studied named as Beyond MSSM (BMSSM) \cite{Dine:2007xi}.
In the above expression, $\Lambda$ is a cutoff scale of the model,
and $c$ is a coupling constant.
Let us redefine the parameter $\Lambda$ to make $c=1$ just to make the expression below simple.
Expanding the SU(2) contract explicitly, we obtain:
\begin{equation}
W = \mu (H_u^+ H_d^- - H_u^0 H_d^0) + \frac{1}{\Lambda} (H_u^+ H_d^- - H_u^0 H_d^0 )^2.
\end{equation}
Then, in the Lagrangian, the charged Higgsino mass and the Higgs to Higgsino coupling
can be extracted as
\begin{equation}
-{\cal L} \supset
\left(\mu  - \frac{2}{\Lambda} H_u^0 H_d^0\right) \tilde H_u^+ \tilde H_d^-.
\end{equation}
The charged Higgs scalar and Higgsino coupling terms are suppressed.
Using 
\begin{eqnarray}
{\rm Re}\, H_d^0 &=& v_d + \frac1{\sqrt2} (H\cos\alpha  - h\sin\alpha ),\\
{\rm Re}\,H_u^0 &=& v_u + \frac1{\sqrt2} (H\sin\alpha  + h\cos\alpha ),
\end{eqnarray}
we obtain the Higgs to Higgsino coupling as
\begin{equation}
g_{hf\bar f} = - \frac{v}{\Lambda} \cos(\alpha+\beta) ,
\end{equation}
where $v/\sqrt2 = \sqrt{v_u^2+v_d^2}$.
If $\mu$ and $\Lambda$ have the same signs,
the Higgsino loop can provide a constructive contribution to the diphoton decay amplitude.
However, in order to obtain a significant contribution,
$\Lambda$ has to be about $v\cos(\alpha+\beta)$,
which means that a large size of non-renormalizable interaction is required.

The non-renormalizable term can induce the Higgs coupling to charged Higgs bosons.
Using
\begin{eqnarray}
H_d^+ &=& \chi^+\cos\beta  - H^+\sin\beta, \\
H_u^+ &=& \chi^+\sin\beta  + H^+\cos\beta,
\end{eqnarray}
where $\chi^+$ is a Goldstone boson eaten by the $W$ boson,
we obtain
\begin{equation}
g_{hH^+H^-} = 
\frac{v\mu}{\Lambda} (\cos(3\beta-\alpha)-3\cos(\alpha+\beta))
-
\frac{v^3}{2\Lambda^2}(\cos(3\beta-\alpha)-5\cos(\alpha+\beta))\sin2\beta.
\end{equation}
The Higgs coupling to vector bosons in two Higgs doublets is given as
$g_{hVV} = g_{hVV}^{\rm SM} \sin(\beta-\alpha)$.
If $h\to ZZ \to 4\ell$ is not reduced
and $W$ loop contribution to diphoton decay width in SM is kept, 
$\beta-\alpha$ should be about 90$^{\rm o}$.
The first term of $g_{hH^+H^-}$ is expected to be larger than the second term because of $\mu \gg v$,
and then, the absolute value of the coupling is maximized when $\beta \sim -\alpha \sim 45^{\rm o}$.
The Higgs coupling with the maximized magnitude is $g_{hH^+H^-} \simeq -4v\mu/\Lambda$,
which is negative if $\mu/\Lambda>0$.

The Higgs mass correction due to the non-renormalizable term is obtained as (see Appendix)\footnote{As one can find from the derivation in Appendix, the 
tree-level Higgs mass does not depend on the soft SUSY breaking mass parameters 
explicitly as long as $Z$ boson mass is fixed. Therefore, we do not mention 
about the size of SUSY breaking to investigate the tree-level corrections to 
Higgs boson mass. As well-known, in MSSM, the minimization condition requires an
 unnatural cancellation between $-m_{H_u}^2$ and $\mu^2$. However, in the model 
beyond MSSM we concern in this paper, the minimization condition is modified, 
and such unnatural cancellation is not required because the Higgs potential is 
lifted by a new term with a scale parameter $\Lambda$. As a consequence, a 
little hierarchy between the SUSY breaking masses and $Z$ boson mass is not very
 unnatural contrary to MSSM.}:
\begin{equation}
\Delta m_h^2 \simeq \frac{3 v^4 \sin^2 2\beta}{\Lambda^2} - \frac{4\mu v^2 \sin2\beta}{\Lambda},
\end{equation}
if the heavier neutral Higgs boson is decoupled.
As explained, in order to obtain a significant enhancement of the diphoton decay rate, 
we require a sizable values of the Higgs couplings,
and as a result, we need a large value of $v/\Lambda$ (for Higgsino loop)
or $\mu/\Lambda$ (for charge Higgs loop).
In the case that Higgsino loop provides a significant contribution,
$v/\Lambda$ has to be very large if $\beta-\alpha \sim 90^{\rm o}$,
and such a large value can induce too large correction to the Higgs mass to obtain 125 GeV mass.
The charged Higgs loop can contribute to the diphoton decay,
and enhance the decay width if $\mu/\Lambda>0$.
However, in that case, it reduces the SM-like Higgs mass.
Although there is freedom to cancel those two corrections by adjusting 
$4\mu/\Lambda = 3 v^2/\Lambda^2 \sin2\beta$,
such cancellation is not very natural.
%
%
%
%
%
%

Totally, the non-renormalizable term is not a good candidate of source to
enhance the diphoton decay width.
It is true that the term can generate the Higgs coupling to charged particles,
but the direction to generate a constructive contribution to the decay amplitude
reduces the Higgs boson mass.
To obtain a significant contribution to the decay width, the size of non-renormalizable term
needs to be large, and it can modify the Higgs boson mass too much.
It may be possible to tune the Higgs mass to be 125 GeV with
enhancing the decay width, but it is not a natural situation.

\section{Non-perturbative Higgs model}

Let us consider the following VEV-dependent fermion mass term:
\begin{equation}
-{\cal L} = \lambda \Lambda \frac{(|H|^2)^a}{\Lambda^{2a} }\bar f f.
\end{equation}
Using $H= v/\sqrt2 + (h+i\phi)/\sqrt2$, we obtain the mass and the Higgs coupling as
\begin{equation}
m_f = \lambda \Lambda \left( \frac{v^2}{2 \Lambda^2} \right)^{a},
\qquad
g_{hf\bar f} = 2 \lambda a \frac{\Lambda}{v} \left( \frac{v^2}{2 \Lambda^2} \right)^{a},
\end{equation}
and
\begin{equation}
\frac{g_{hf\bar f}}{m_f} = \frac{2a}{v}.
\end{equation}
Therefore, as far as we consider the perturbative term (i.e. $a$ is a natural number),
the fermion loop provides a destructive contribution to the diphoton decay amplitude in SM.
However, in non-perturbative case, the exponent $a$ can be negative,
and the fermion loop enhances the diphoton decay width. 
In fact, we know an example of negative $a$ in SUSY QCD (SQCD)
as a runaway potential generated by instanton effects \cite{Affleck:1983rr}.

In $SU(N)$ SQCD with $N_f$ flavor, the runaway non-perturbative potential
is generated if $N>N_f$.
The representations of matter chiral superfields
under the symmetry $SU(N) \times SU(N_f)_L \times SU(N_f)_R$ are
\begin{equation}
Q : ({\bf N},{\bf N_f},1), \quad \bar Q (\bar {\bf N}, 1, {\bf N_f}) .
\end{equation}
The non-perturbative superpotential is
\begin{equation}
W \propto \frac{\Lambda^{3+ \frac{2N_f}{N-N_f}}}{(\det \bar Q Q)^{\frac{1}{N-N_f}}}.
\end{equation}

In order to construct a Higgs model,
let us consider the case of $N_f=2$.
Suppose that $SU(N_f)_L$ is the weak gauge symmetry,
and a $U(1)$ subgroup of $SU(N_f)_R$ is the hyper charge symmetry.
(In this case, the color number $N$ of SQCD should be even to eliminate $SU(2)_L$ anomaly).
Moduli fields of SQCD, $\bar Q Q$, can be identified as a Higgs bidoublet:
\begin{equation}
\Lambda H_1^a = \bar Q_1 Q^a, 
\quad
\Lambda H_2^a = \bar Q_2 Q^a.
\end{equation}
One can easily find $\det \bar Q Q = \Lambda^2 H_1 \cdot H_2$ (where $\cdot$ stands for
a SU(2) contract),
and therefore, we obtain
\begin{equation}
W = c \frac{\Lambda^{3+2\kappa}}{(H_1 \cdot H_2)^\kappa},
\end{equation}
where $\kappa = 1/(N-2)$. This kind of superpotential for the composite Higgs model has been considered in \cite{Haba:2004bz}. Redefining $\Lambda$, we will choose $c=1$ hereafter.

As it is called as runaway potential,
if there is non-perturbative potential alone,
the vacua goes to infinity along the flat direction.
However, if the scalar potential is lifted due to SUSY breaking terms, 
the potential can be stabilized 
and the chiral symmetry is spontaneously broken \cite{D'Hoker:1996cs}.
As a result, Higgsino mass is generated non-perturbatively,
and its scale is determined by the non-perturbative scale $\Lambda$
and the SUSY breaking scale.
Therefore, this model can be one of the solution of so-called $\mu$-problem
(i.e. origin of the Higgsino mass).

The interesting point in this model is that
the Higgsino loop generates a constructive contribution
to the diphoton decay amplitude
due to the fact that the fields are placed in the denominator.

Let us describe the Higgsino mass 
and the Higgs coupling from the superpotential:
\begin{equation}
W = \frac{\Lambda^{3+2\kappa}}{(H_u^0 H_d^0 - H_u^+ H_d^-)^{\kappa}}.
\end{equation}
The K\"ahler metric of the Higgs fields may not be canonical.
However, we assume the canonical form of the K\"ahler metric (just for simplicity),
and we neglect terms from K\"ahler connection.
Suppressing the charged Higgs scalar and Higgsino coupling terms,
we obtain 
\begin{eqnarray}
-{\cal L} &\supset&
-\kappa \Lambda^{3+2\kappa} (H_u^0 H_d^0)^{-\kappa-1} \tilde H_u^+ \tilde H_d^- \\
&= &
-\kappa \Lambda \left(\frac{\Lambda^2}{v_u v_d}\right)^{1+\kappa}
\left(1- (\kappa+1) \frac{2}{v} \frac{\cos(\alpha+\beta)}{\sin2\beta} h + \cdots\right)  \tilde H_u^+ \tilde H_d^- .
\end{eqnarray}
and
\begin{equation}
\frac{g_{h \tilde H^+ \tilde H^-}}{m_{\tilde H^+}} = 
-(\kappa+1) \frac{2}{v} \frac{\cos(\alpha+\beta)}{\sin2\beta}. 
\end{equation}
Note that 
$\frac{\cos(\alpha+\beta)}{\sin2\beta} \simeq 1$
if $h$ is the SM-like Higgs (i.e. $\sin(\beta-\alpha)\simeq 1$)
and the heavier Higgs mass is much larger than the $Z$ boson mass (i.e. $\tan 2\alpha \simeq \tan2\beta$). 
Therefore, it can easily generate the constructive contribution to the 
diphoton decay.

The ratio of the decay amplitude from top quark loop and Higgsino loop can be obtained as
\begin{equation}
\frac{A^{\gamma\gamma}_{\tilde H^+}}{A^{\gamma\gamma}_t}
\simeq
-\frac{2(1+\kappa)}{3 (2/3)^2} \frac{A_{1/2}(\tau_{\tilde H^+})}{A_{1/2}(\tau_{t})}
\simeq -1.5 \times (1+\kappa).
\end{equation}
If the charged Higgsino is heavier than 100 GeV, $A_{1/2}$ loop function does not 
have much difference between top and Higgsino loops.
The decay amplitude depends on $\kappa = 1/(N-2)$,
and $N$ is a number of color in SQCD.
We find that the Higgs to Higgsino coupling can modify the
SM amplitude 
about at least 40\% ($1.5 \times 1.84/(8.32-1.84) = 0.42$),
and it can enhance the decay width twice as the SM one.
Interestingly,
the recent results at the ATLAS experiments,
the center value of the diphoton rate
is about twice as large as the SM prediction.

The scalar potential from the superpotential is obtained as
\begin{equation}
V = \Lambda^{6+4\kappa}
\frac{|H_u^0|^2+|H_u^+|^2 + |H_d^0|^2 + |H_d^-|^2 }{|H_u^0 H_d^0 - H_u^+ H_d^-|^{2(\kappa+1)}}.
\end{equation}
The Higgs coupling to the charged Higgs from the non-perturbative scalar potential is
\begin{equation}
g_{h H^+ H^-} = 4(1+\kappa) 
\left(
\frac{\Lambda^2}{v_u v_d}
\right)^{2(\kappa+1)}
\frac{\Lambda^2}{v} \left(
2(1+\kappa) \frac{\cos(\alpha+\beta)}{\sin2\beta}- \sin(\beta-\alpha)
\right).
\end{equation}
The Higgs coupling to the charged Higgs bosons is positive
for $\cos(\alpha+\beta)/\sin2\beta \simeq 1$ and $\sin(\beta-\alpha)\simeq 1$, 
and it induces a destructive contribution
to the diphoton decay amplitude.
The charged Higgs mass can (mainly) come from SUSY breaking mass terms,
and therefore, $g_{hH^+H^-}/m_{H^+}^2$ depends on the charged Higgs mass
and is less predictive than the Higgsino case.
If the charged Higgs is much heavier than the lightest Higgs boson,
the contribution is not significant.

The neutral Higgs mass correction from the $F$-term scalar potential is obtained
as
\begin{equation}
\Delta m_h^2 
= 4 (1+\kappa)(1+2\kappa) \Lambda^2 
\left( 
\frac{\Lambda^2}{v_u v_d}
\right)^{2(1+\kappa)}.
\end{equation}
We note that 
the correction of the Higgs boson mass
is comparable to the Higgsino mass.
Therefore, if there are no additional non-quadratic terms
(quadratic terms are consumed to satisfy the stationary condition
and those freedom is fixed by the VEVs of Higgs fields,
and so, Higgs mass does not depend on the quadratic terms explicitly),
the naive size of the Higgsino mass is about 100 GeV.
However, such light Higgsino with a sizable coupling to Higgs
is harmful phenomenologically,
because the Higgsinos can be resonantly produced by the Higgs coupling to Higgsinos.
At least, the neutral Higgsino (which can also have a large Higgs coupling) 
should not be the lightest SUSY particle
(if $R$-parity conservation is assumed) since no missing energy is observed.
Because of the non-perturbative effects,
other types of SUSY breaking term can be generated,
and it may break the naive relation between Higgs boson and Higgsino masses.
One can add a VEV-independent Higgsino mass to avoid a possible difficulty,
though the predictivity and the motivation to solve $\mu$ problem are lost.
We do not go to the detail of the particle spectroscopy in this paper.
The particle spectroscopy and the phenomenological study of the non-perturbative Higgs model
will be studied somewhere else \cite{progress}.

Before concluding this section, 
we comment on the Yukawa interaction to SM fermions.
Possible Yukawa interaction to top quark is generated by a non-renormalizable term
\begin{eqnarray}
W_Y = \frac{1}{M_*} q_L t_R^c Q \bar Q_2,
\end{eqnarray}
where $q_L$ is a left-handed quark doublet and $t_R^c$
is a right-handed quark field.
As described in this section,
the (up-type) Higgs superfield is a composite of the SQCD fields $Q$, $\bar Q$ :
$H_u = Q\bar Q/\Lambda$. 
A proper top mass requires $M_* \sim \Lambda$.
In order to obtain a proper size of top quark mass,
one can also consider an extension of the SQCD model
in which top quark fields are also moduli fields in SQCD.
For example,
the number of SQCD flavor is chosen as $N_f = 6$,
and the SM gauge group $SU(3)_c \times SU(2)_L \times U(1)_Y$
is a subgroup of $SU(6)_L \times SU(6)_R$.
In order to describe it simply, let us use Pati-Salam symmetry base:
$SU(4)_c \times SU(2)_L \times SU(2)_R$
(The gauged symmetry can be the SM subgroup of the Pati-Salam symmetry).
$SU(4)\times SU(2) \times U(1)$ is a subgroup of $SU(6)$,
and diagonal subgroup of $SU(4)_L \times SU(4)_R$
is $SU(4)_c$.
The composite field $\bar Q Q$ can be decomposed as
\begin{equation}
\left(
\begin{array}{cc}
H:({\bf1},{\bf2},{\bf2}) & L:({\bf 4},{\bf 2},{\bf 1}) \\
\bar R:(\bar{\bf4},{\bf1},{\bf2}) & 
\Sigma:({\bf 15},{\bf 1},{\bf 1}) + S: ({\bf 1},{\bf 1},{\bf 1})
\end{array}
\right).
\end{equation}
The third generation fields (top, bottom and tau) are unified 
in the SQCD moduli fields $L,\bar R$.
The non-perturbative superpotential is
$W_{np} \propto 1/({\rm det}\, \bar Q Q)^\alpha$,
and $\det \bar Q Q = (S+\Sigma)^4 H_1 H_2 + L \bar R H (S+\Sigma)^3 +
LL\bar R \bar R (S+\Sigma)^2 + \cdots$.
The fermion masses of the third generation are obtained by
$\frac{\partial^2 W_{np}}{\partial L \partial \bar R}$,
and the order 1 size of (effective) Yukawa coupling can be generated
naturally.
In this setup, the Higgs couplings to fermions can be different from the SM ones.
Actually, the signature of the Higgs coupling to top quarks can be opposite to SM
in the same way as Higgsino case, and
top quark contribution can become constructive to the W boson loop.
In this kind of model,
the Higgs coupling to SM fermions may be different from SM and
further investigation is needed. 
The LHC experiments have not yet observed $h \to \tau\tau$
and $h \to bb$ via Yukawa couplings
(though $h\to bb$ is indicated by Tevatron 
\cite{:2012cn})\footnote{The CMS and ATLAS experiments are 
starting to observe these channels but there is still room for considerations of new physics~\cite{HCP}.}.

\section{Conclusion and Discussion}

The observation of a new boson which is consistent with SM
opens a new era of Higgs physics.
It is important to investigate 
all the decay modes
to see that the new boson is really identified to the 
SM Higgs boson.
Among the decay modes,
a possible hint beyond SM is the enhancement of diphoton decay rate
indicated by ATLAS/CMS measurements.
The diphoton decay rate is larger than the SM expectation
since the 2011 data,
though there is no enough statistics yet.
The Higgs to diphoton decay
is induced at the loop level,
and therefore, it is sensitive to new physics beyond SM.
Motivated from the excess of diphoton decay rate,
we investigate the possibility
if the loop diagram via Higgsino (which is a superpartner of the Higgs boson)
can enhance the diphoton decay width.

We first analyze an additional non-renormalizable term of Higgs fields.
The higher order term can introduce the Higgs coupling to Higgsino
as well as charged Higgs bosons.
In this type of model, however,
we learn that
the choice of the Higgs coupling to obtain a significant size of constructive contribution
to the diphoton decay amplitude in SM via $W$ boson loop 
can reduce the Higgs boson mass (which is not preferable to obtain 125 GeV Higgs boson mass
since the MSSM Higgs boson mass is less than $Z$ boson mass at the tree-level)
and/or the size of non-renormalizable term needs to be large
and a cutoff scale may be just around the weak scale.
There may be a solution in a complicate situation, 
but it is not quite attractive.

We suggest another model in which
the Higgsino mass term is generated by a non-perturbative instanton effect
via a strong dynamics in SUSY QCD.
In this kind of model,
the bosonic fields are in the denominator of the operator,
and thus,
the Higgs coupling to fermions flips its sign
compared to the perturbative (polynomial) interaction.
As a result, the constructive contribution to the diphoton decay amplitude
is easily obtained.
If the Higgsino mass purely comes from the non-perturbative superpotential,
the loop correction of the diphoton decay amplitude is predictive,
and the Higgsino loop can enhance the decay width
(more than) twice as large as SM prediction.
Interestingly,
the current center value of the diphoton decay rate is
about two times larger than the SM expectation.

Various types of non-perturbative Higgs coupling
can be constructed using the strong dynamics in SQCD.
In those models,
the Higgs couplings to the quarks and leptons 
can be also modified from the SM ones.
The LHC experiments will soon provide 
tons of data to see various decay modes,
%
and the non-perturbative Higgs model can be tested.

\appendix

\section{Appendix: Higgs mass}

In this appendix,
we show how to obtain the (tree-level) physical Higgs mass
in a general form scalar potential.

In the beginning,
let us describe a case of single Higgs doublet $H$.
The scalar potential is in general a function of $|H|^2$:
\begin{equation}
V = V(|H|^2).
\end{equation}
Denoting
\begin{equation}
H = \left(
 \begin{array}{c}
  \chi^+ \\
  \frac{v}{\sqrt2} + \frac{h + i\chi}{\sqrt2}
 \end{array}
\right),
\end{equation}
we obtain
\begin{equation}
|H|^2 = \frac{v^2}{2} + v h + \frac{h^2+ \chi^2}{2} + \chi^+ \chi^-.
\end{equation}
Expanding the potential around the VEV $v$, we obtain
\begin{equation}
V = V (v^2/2)+ V^\prime (v^2/2)(v h + \frac{h^2+ \chi^2}{2} + \chi^+ \chi^-)
+ \frac12 V^{\prime\prime} (v^2/2) (v h + \frac{h^2+ \chi^2}{2} + \chi^+ \chi^-)^2.
\end{equation}
The stationary condition (vanishing the linear term of $h$) is $V^\prime = 0$.
Then, one can find that $\chi$ and $\chi^+$ are massless Goldstone bosons,
and would be eaten by the gauge bosons.
The mass of the physical Higgs boson $h$ is easily obtained
as $m_h^2 = v^2 V^{\prime\prime}$.
For example, in $\phi^4$ theory,
$V(x) = m^2 x + \lambda x^2$,
and we obtain
$m_h^2 = 2\lambda v^2$.

In the case of two Higgs doublet $H_1$, $H_2$,
the scalar potential is a function 
of $|H_1|^2$, $|H_2|^2$ and $H_1 \cdot H_2$. 
In order to make the following calculation simple,
it is convenient to define linear combinations of the Higgs doublet:
\begin{equation}
\Phi_1 = H_1 \cos\beta + \hat H_2 \sin\beta ,
\quad
\Phi_2 = -H_1 \sin\beta + \hat H_2 \cos\beta,
\end{equation}
where $\hat H = i\sigma_2 H^*$,
so that the VEV of $\Phi_2^0$ is zero by definition.
We define 
\begin{equation}
x = |\Phi_1|^2, \quad
y = |\Phi_2|^2, \quad
z = \Phi_1 \cdot \hat\Phi_2 , \quad
\bar z = \Phi_2 \cdot \hat \Phi_1,
\end{equation}
and the general potential is a function $V(x,y,z,\bar z)$.
The stationary conditions are
$V_x = V_z = V_{\bar z} = 0$,
where $V_x$ denotes a partial derivative by $x$ for example.

Expanding the potential
around the VEV, $\langle x \rangle = v^2/2$,
we obtain the mass term of the neutral Higgs bosons:
\begin{equation}
\frac12
(\begin{array}{cc}\Phi_1^0 & \Phi_2^0 \end{array})
\left(
\begin{array}{cc}
v^2 V_{xx} & \frac{v^2}2 (V_{xz}+ V_{x\bar z}) \\
\frac{v^2}2 (V_{x z} + V_{x\bar z}) & V_y + \frac14 v^2 (V_{zz} + V_{\bar z\bar z} + 2 V_{z\bar z})
\end{array}
\right)
\left(\begin{array}{c}\Phi_1^0 \\ \Phi_2^0 \end{array}\right).
\end{equation}
If $V_y$ is large and $\Phi_1^0$-$\Phi_2^0$ mixing is small,
$\Phi_1^0$ is roughly the lightest Higgs boson,
and 
$m_h^2 \simeq v^2 V_{xx}$.
The mass of CP odd Higgs boson $A$ is obtained as 
$m_A^2 = V_y  + \frac14 v^2 (-V_{zz} - V_{\bar z\bar z} + 2 V_{z\bar z})$.
The charged Higgs mass is $m_{H^+}^2 = V_y$.

The Higgs mass corrections from the additional potential in the text are obtained
by $v^2 V_{xx}$
by using the following expressions :
\begin{eqnarray}
|H_1|^2 &=& x \cos^2\beta + y\sin^2\beta - \frac12 (z+\bar z) \sin2\beta, \\
|H_2|^2 &=& x \sin^2\beta + y\cos^2\beta + \frac12 (z+\bar z) \sin2\beta, \\
H_1 \cdot H_2 &=& \frac12 (x-y) \sin2\beta + z \cos^2\beta - \bar z \sin^2\beta.
\end{eqnarray}

We exhibit examples of the Higgs mass calculations using the general expressions above.
First, we consider the case of MSSM.
The $D$-term scalar potential in terms of $|H_1|^2$, $|H_2|^2$ and $H_1 \cdot H_2$ is 
$V_D= (g^2+g^{\prime2})/8 (|H_1|^2-|H_2|^2)^2+g^2/2(|H_1|^2|H_2|^2-|H_1\cdot H_2|^2)$.
Extracting the neutral component of the potential (in order to exhibit the essential part), 
we obtain the quartic Higgs coupling as
$g_Z^2/8 (|H_1^0|^2 - |H_2^0|^2)^2 = g_Z^2/8 (x-y)^2 \cos^22\beta$, where $g_Z^2 = g^2+g^{\prime 2}$.
Because $V_{xx}|_{y=0}= g_Z^2/4 \cos^22\beta$,
 the lightest Higgs mass is roughly obtained as 
$m^2_h \simeq g_Z^2/4 v^2 \cos^22\beta = M_Z^2 \cos^22\beta$
at the tree-level.

Secondly, let us derive the Higgs mass corrections from simple extensions of MSSM.
In the case of Next-to-MSSM, $\lambda S H_1 \cdot H_2$ term in the superpotential
generates a quartic Higgs coupling
$\lambda^2 |H_1 \cdot H_2|^2 
=\lambda^2 |1/2 (x-y)\sin2\beta + z \cos^2\beta- \bar z\sin^2\beta|^2$.
One obtains $\partial^2 |H_1\cdot H_2|^2/\partial x^2|_{y=z=\bar z=0}= 1/2 \sin^22\beta$,
and $\Delta m_h^2 \simeq 
v^2 \lambda^2 \partial^2 |H_1\cdot H_2|^2/\partial x^2 = \lambda^2 v^2/2 \sin^22\beta$.
For small $\tan\beta$, this contribution is helpful
to enlarge the lightest Higgs mass as it is well known.
One can add 
an SU(2) adjoint $\Sigma$ (with hypercharge $Y= -1$) and a superpotential term
$\lambda_2 H_2^\alpha H_2^\beta \Sigma_{\alpha \beta}$ ($\alpha$ and $\beta$ are
SU(2) indices).
Then, additional potential
$\lambda_2^2 |H_2|^4$ is generated,
and 
we obtain $\Delta m_h^2 \simeq 
2 \lambda_2^2 v^2 \sin^4\beta$.
This term is helpful to enlarge the Higgs mass in the case of large $\tan\beta$.





\section*{Acknowledgments}

This work is partially supported by Scientific Grant by Ministry of 
 Education and Science, Nos. 22011005, 24540272, 20244028, and 21244036.
The work of Y.M. is supported by the Excellent Research Projects of
 National Taiwan University under grant number NTU-98R0526. The works of K.K. and R.T. are supported by Research Fellowships of the Japan Society for the Promotion of Science for Young Scientists.


\end{document}